\documentclass[epj]{svjour}
\usepackage{amsmath}
\usepackage{amsfonts}
\usepackage{graphicx}
\usepackage{appendix}
\usepackage{dsfont}
\usepackage{amssymb}
\usepackage{bm}
\usepackage{color}
\usepackage{ulem}
\usepackage{soul}
\usepackage{natbib}
\setstcolor{red}
\usepackage{nccmath}
\usepackage{array}

\newcommand{\beq}{\begin{equation}}
\newcommand{\eneq}{\end{equation}}
\begin{document}
\title{Junction of three off-critical quantum Ising chains and two-channel Kondo effect 
in a superconductor}
\author{Domenico Giuliano\inst{1,2}, Gabriele Campagnano\inst{3}, and Arturo Tagliacozzo\inst{3,2}}

%
\institute{Dipartimento di Fisica, Universit\`a della Calabria Arcavacata di Rende I-87036, Cosenza, Italy
\and I.N.F.N., Gruppo collegato di Cosenza, Arcavacata di Rende I-87036, 
Cosenza, Italy 
\and  Dipartimento di Fisica, Universita’ di Napoli ``Federico II''
and CNR-SPIN, Monte S. Angelo-Via Cintia, I-80126, Napoli, Italy}
\date{Received: date / Revised version: date}
%
\abstract{
We show that  a junction  of three off-critical quantum Ising chains can be  
regarded as a quantum spin chain realization 
of the two-channel spin-1/2 overscreened Kondo  effect with two superconducting leads. 
We prove that, as long as the Kondo temperature is larger than the superconducting gap,
the equivalent Kondo model flows towards the  2 channel Kondo fixed point. 
We argue that  our 
system provides the first controlled realization of 2 channel Kondo effect with superconducting 
leads. Besides its theoretical interest, this result is of  importance for  
potential applications to a number of contexts, including the analysis of the   
quantum entanglement properties of a Kondo system.
  } 

\PACS{ 	
      {75.10.Pq}{Spin chain models}   \and
      {72.10.Fk}{Scattering by point defects, 
      dislocations, surfaces, and other imperfections 
      (including Kondo effect)} \and
      {71.10.Pm}{Fermions in reduced dimensions 
      (anyons, composite fermions, Luttinger liquid, etc.) }
     } 
\titlerunning{Junction of 3 quantum Ising chains and 2-channel Kondo effect in a superconductor}
\authorrunning{D. Giuliano et al.}
\maketitle
\section{Introduction}
\label{intro}

The Kondo effect \cite{hewson_book} and the superconductivity \cite{tinkham_book} are among the  most
remarkable effects of many-body correlations in condensed matter systems. Specifically, 
in the former case itinerant electrons conspire to screen a localized magnetic impurity in 
a conducting media to an isolated spin singlet; in the latter
case, electrons pair in two-particle Copper pairs, and eventually condense in a collective
ordered state, in which single-particle excitations are fully gapped, with the dependence of 
the gap on the momentum determined by the specific superconducting state which is created. 
Remakably, the simultaneous presence of the two effects gives rise to an interesting 
competition: indeed, it is well-known that Kondo effect is strictly related to
low-energy singularities in the single-fermion scattering amplitude of the magnetic 
impurities, close to the Fermi surface. This clearly conflicts with the presence of an energy gap 
in the single-fermion spectrum in the superconducting phase, which makes the density of 
states in the vicinity of the Fermi level equal to 0. Nevertheless, 
despite the  gap, Kondo effect is not necessarily 
suppressed by the onset of superconductivity. This is due to the well-known
result that the fermions effectively
screening the impurity are the ones at energies (measured with 
respect to the Fermi level) ranging from the 
half-bandwidth all the way down to   $k_B T_K$, with $T_K$ being  the Kondo temperature
and $k_B$ the Boltzmann constant 
\cite{anderson}. Therefore, Kondo effect is expected to persist even in a superconducting 
medium with gap $\Delta$, provided  $k_B T_K \gg \Delta$, which makes the gap itself 
immaterial for the screening of the magnetic impurity \cite{buitelaar}.

In the last two decades, the enormous progress in the fabrication techniques of nanostructures 
made it possible to realize Kondo effect  in a controlled way in e.g. quantum dots at Coulomb 
blockade \cite{kouwenhove_1,goldhaber_gordon_1} or in single magnetic 
impurities \cite{procolo} in  contact with metallic leads. More generally,
Kondo effect in low-dimensional systems has been explored \cite{luca_dellanna}, especially in view of 
its relation to remarkable many-body collective effects \cite{avishai_book}, such as, for instance  the electronic shake-up 
after a single-electron emission \cite{ort_kondo,sindona_1}.
This motivated further proposals for 
studying the coexistence/competition between Kondo effect and superconductivity 
in a quantum dot coupled to superconducting leads, also in a Josephson-junction arrangement
(dot connected to two superconducting leads at zero voltage bias and 
fixed phase difference), which 
should be able to evidence the crossover between $\pi$-junction (no Kondo effect) 
and 0-junction (onset of the Kondo effect) \cite{avishai,choi_1,choi_2,campagnano_1}. 
On the experimental side, a remarkable  scaling law of 
the dc-conductance, which results to be a universal function of $ \Delta / ( k_B T_K )$,
has been observed   in a  single  
quantum dot contacted laterally to a superconducting reservoir \cite{tarucha}. 
While there is still  some debate about nonuniversal features, it is basically 
estabilished that the onset of Kondo effect 
is effective whenever $K_B T_K / \Delta \gg 1$ and that the fixed point as $ T \to 0$ should correspond 
to the perfectly screened Nozi\`eres fermi liquid \cite{hewson_book}. Other issues which have been
studied using quantum dots connected to superconductors are, for instance, the interplay between Kondo effect
and Andreev reflection in dots coupled to one normal and one superconducting lead \cite{tanaka}, or in 
a dot coupled to topological superconducting leads \cite{lopez}.

Recently, novel possible realizations of Kondo effect have been proposed at 
junctions of interacting quantum wires and topological superconductors 
\cite{beri_cooper,egger_1,egger_2,egger_3}, in an SNS-junction made with 
topological superconductors (where it should be detected by looking at the scaling 
of the current with the system size) \cite{GiuAf2,GiuAf3}, or in junctions of  quantum spin chains 
\cite{crampe_trombettoni,tsvelik_Ising,giuliano_sodano_epl,gstt_1}. In particular, 
the quantum spin chain 
realization of the Kondo effect presents a number of theoretically interesting features, such 
as the possibility of realizing in a ''natural'' way the symmetry between channels in the
many-channel version of the effect \cite{tsvelik_Ising} or, on the theoretical side, the 
exact integrability of some specific models \cite{tsvelik_exact,buccheri}. On the 
applicative side, it appears particularly intriguing, due to the possibility of 
realizing  in a controlled way devices behaving as spin chains and/or as junctions of
spin chains by means, for instance, of pertinently engineered superconducting 
quantum wires \cite{giu_so_v0}, or of  quantum Josephson junction networks 
\cite{giu_so_v3,giu_so_v2,giu_so_v1}.  

In a junction of quantum spin chains the magnetic impurity is determined by 
the coupling between the chains. Formally, this is evidenced by extending to 
the junction the  Jordan-Wigner transformation \cite{jordan_wigner}, by means of 
which one realizes quantum spin-1/2 operators in terms of lattice spinless fermion 
operators (and vice versa). When applied to a junction of more than two chains, the 
Jordan-Wigner tranformation requires introducing ancillary fermionic degrees of 
freedom, to preserve the correct commutation relations between corresponding 
operators. At a three-chain junction, 
this determines an effective spin-1/2 magnetic impurity which is 
topological, due to the nonlocal character of the ancillary degrees of 
freedom \cite{crampe_trombettoni}. 
The Jordan-Wigner fermion can, therefore, act to realize Kondo effect by 
screening this effective magnetic impurity. 
Along this correspondence,  
in order to recover a gapless single-fermion spectrum, an important requisite is 
that the chains forming the junction are all tuned at a quantum critical point, 
either corresponding to the paramagnetic-ferromagnetic phase transition in 
the quantum-Ising chains  \cite{tsvelik_Ising,giuliano_sodano_epl} or in the 
XY quantum spin chains \cite{gstt_1}, or belonging to a critical line of gapless
points, such as in the junction of quantum XX spin chains \cite{crampe_trombettoni}.

In this paper we rather focus onto the Kondo effect at a junction of 
three off-critical (on either the paramagnetic, or the ferromagnetic
side) quantum Ising chains, with a nonzero gap in the single-fermion spectrum.
Specifically, by going through a rigorous mapping between the off-critical junction 
of Ising chains and the model for a spin-1/2 magnetic impurity interacting with 
two  superconducting baths, we prove that our system can be regarded as a model
for two-channel Kondo effect with superconducting leads. 

A first important feature of the system we consider is that it hosts a remarkable
realization of overscreened, spin-1/2 two-channel Kondo effect with 
superconducting electronic baths
which, so far, has never emerged in realistic devices based on e.g. quantum dots with 
superconducting leads. Moreover, our system  naturally presents a 
symmetry between channel, which is typically hard to recover in ''standard'' condensed
matter-based  many-channel Kondo systems \cite{tsvelik_Ising,tsvelik_mapping}. Finally, the very 
fact that our system is based on a junction of spin chains makes it possible to use it 
for potentially countless numerically- and analitically-based 
applications, such as, for instance, probing the effects of superconductivity 
on the entanglement structure of the system \cite{sodano_et_al} and, more generally, 
verifying how the Kondo interaction affects the 
entanglement of the spin chains close to their ''bulk'' quantum critical point
\cite{amico,sindona}.

The paper is organized as follows:

\begin{itemize}
 
 \item In section \ref{modhamju}, we introduce the model Hamiltonian for the junction of 
 three quantum spin chains and map it onto a pertinent fermionic Hamiltonian by employing an 
 adapted version \cite{crampe_trombettoni} of the Jordan-Wigner transformation;
 
\item In section \ref{mapping}, we rigorously trace out the mapping between the Jordan-Wigner fermionic
representation of the junction Hamiltonian and a model for a quantum spin-1/2 impurity interacting with 
two superconducting baths (''channels'');

\item In section \ref{per_kondo}, we analyze the onset of Kondo regime by means of a pertinently
adapted version \cite{gstt_1} of poor man's renormalization group approach to Kondo problem \cite{anderson}, 
finding the necessary conditions to which the Kondo coupling and the single-fermion energy gap must obey, 
in order to actually recover the Kondo effect;

\item In section \ref{majo_kondo}, we discuss whether, and how,  Majorana-fermion-like excitations arising at 
the endpoints of the chains in the magnetically ordered phase affect the  Kondo effect, 
proving that their are basically irrelevant for what concerns Kondo physics;

\item In section \ref{strocou}, we describe the strongly coupled Kondo fixed point of the system using a variational 
approach \cite{campagnano_1,giuliano_tagliacozzo_1}, adapted to the specific case of gapped leads.
We conclude that the gap does not substantially affect the structure of the Kondo fixed point, provided
the conditions for the onset of Kondo regime are met;

\item In section \ref{conclusions}, we provide our main conclusions, together with a discussion of 
possible further developments of our work;

\item In appendix \ref{appe_a}, we present  mathematical details about the exact solution of a quantum
Ising chain with open boundary conditions in terms of  Jordan-Wigner fermions.
 
\end{itemize}

\section{The model Hamiltonian for the junction}
\label{modhamju}

The possibility of realizing two-channel Kondo (2CK)  effect at a junction 
of three   critical  ferromagnetic quantum Ising chains (QIC)s  
was  originally  put forward by Tsvelik \cite{tsvelik_Ising}  who, 
later on, also proved the exact solvability of the model, taken in 
the continuum limit \cite{tsvelik_exact}. Here, we consider the 
generic situation of a junction of three, non (necessarily) critical
QICs.  Following Tsvelik's construction, 
 we   focus onto a junction of three equal chains,
each one consisting of $\ell$ sites. The three (disconnected) chains are described 
by the model Hamiltonian 

\beq
H_{\rm Chain} = \sum_{ \lambda = 1}^3 \: \Biggl\{ - J \sum_{ j = 1}^{ \ell -1  } 
S_{ j + 1 , \lambda}^x S_{ j , \lambda}^x  +  h \sum_{ j = 1}^\ell S_{ j , \lambda}^z
\Biggr\}
\:\:\:\: . 
\label{mh.1}
\eneq
\noindent
In Eq.(\ref{mh.1}), $S_{ j , \lambda}^x$ and  $S_{ j , \lambda}^z$  are quantum, 
spin-1/2 operators acting on site-$j$ of chain-$\lambda$, $J$ ($>0$) is the ferromagnetic
exchange strength between spins on nearest neighboring sites, $h$ is the applied 
magnetic field in the $z$-direction. With the normalization we chose in 
Eq.(\ref{mh.1}), the chains become quantum critical at $J = \pm h /2$ \cite{sachdev_book}. 
The junction is constructed by connecting the three spins at the endpoints of the 
three chains by means of a ferromagnetic coupling $J_\Delta < J$.
The corresponding boundary Hamiltonian is 
therefore given by 

\beq
H_\Delta = - J_\Delta \sum_{ \lambda = 1}^3  S_{ 1 , \lambda + 1 }^x S_{ 1 , \lambda}^x
\:\:\:\: , 
\label{mh.2}
\eneq
\noindent
with periodicity in the index $\lambda$ understood, that is, $S_{ 1 , \lambda + 3}^x = 
S_{ 1 , \lambda  }^x $. The whole system is described 
by the model Hamiltonian $H = H_{\rm Chains} + H_\Delta$. The mapping of the spin-chain 
junction onto a fermionic Kondo-like Hamiltonian is based onto a generalization 
of the Jordan-Wigner (JW) fermionization procedure for a single chain with open boundary 
conditions, which we review in appendix \ref{appe_a}. Specifically, in order to 
preserve the correct (anti)commutation relations between operators acting on 
different chains, one has to introduce a set of Jordan-Wigner spinless lattice
fermions per each chain, $\{ a_{ j , \lambda }, a_{ j , \lambda}^\dagger \}$, 
in analogy to what is tipically done for a single chain (see appendix \ref{appe_a}
for details) and, in addition, three real-fermionic Klein factors (KF)'s $\sigma^\lambda$, 
one per each  chain \cite{crampe_trombettoni}.   By definition, each $\sigma^\lambda$ 
anticommutes with all the  $a_{ j , \lambda'} ,a_{ j , \lambda'}^\dagger$. On introducing the KFs, 
the JW transformations in Eq.(\ref{app.2})
of appendix \ref{appe_a} are generalized to \cite{crampe_trombettoni,tsvelik_Ising,gstt_1}

\begin{eqnarray}
S_{j , \lambda}^+ &=& i a_{j , \lambda}^\dagger \: e^{ i \pi \sum_{ r = 1}^{ j  - 1 } 
a_{r , \lambda}^\dagger a_{r , \lambda} } \sigma^\lambda  \nonumber \\
S_{j , \lambda}^- &=& i a_{j , \lambda}   \: e^{ i \pi \sum_{ r = 1}^{ j - 1 } 
a_{r , \lambda}^\dagger a_{r , \lambda} }  \sigma^\lambda \nonumber \\
S_{j , \lambda}^z &=& a_{j , \lambda}^\dagger a_{j , \lambda} - \frac{1}{2}
\:\:\:\: . 
\label{mh.3}
\end{eqnarray}
\noindent
Due to the identity $ ( \sigma^\lambda )^2 = 1$, it is easy to check that, when 
inserting Eqs.(\ref{mh.3}) into Eq.(\ref{mh.1}), the KFs fully disappear from 
$H_{\rm Chain}$ and that, accordingly, one obtains 

\begin{eqnarray}
H_{\rm Chain} &=& \sum_{ \lambda = 1}^3 \Biggl\{ 
- \frac{J}{4}\sum_{ j = 1}^{ \ell - 1 } \:
\{ a_{ j , \lambda}^\dagger a_{ j + 1 , \lambda} + 
a_{ j + 1 , \lambda}^\dagger a_{ j , \lambda} \} \nonumber \\
&-& \frac{J}{4} \sum_{ j = 1}^{ \ell - 1 } \:
\{ a_{ j , \lambda} a_{ j + 1 , \lambda} + 
a_{ j + 1 , \lambda}^\dagger a_{ j , \lambda}^\dagger \} + h  \sum_{ j = 1}^{ \ell  } 
 a_{ j , \lambda}^\dagger a_{ j   , \lambda}  \Biggr\}
 \:\:\:\: . 
 \label{mh.4}
 \end{eqnarray}
\noindent
At variance, the KFs do explicitly appear in $H_\Delta$, which  takes the form

\beq
H_\Delta = \sum_{ \lambda = 1}^3 {\cal T}^\lambda \Sigma_1^\lambda 
\:\:\:\: , 
\label{mh.5}
\eneq
\noindent
with 

\beq
\Sigma_j^\lambda = - \frac{i}{2} \: \sum_{ \lambda' , \lambda^{''}} \: 
\epsilon^{ \lambda , \lambda' , \lambda^{''}} 
\: [ a_{ j , \lambda'}^\dagger + a_{ j , \lambda' } ] 
[ a_{ j , \lambda^{''}}^\dagger + a_{ j , \lambda^{''}} ] 
\:\:\:\: , 
\label{mh.6}
\eneq
\noindent
and the effective spin-1/2 operator $\vec{\cal T}$ being given by 

\beq
{\cal T}^\lambda = - \frac{i}{2} \: \sum_{ \lambda' , \lambda^{''}} \: 
\epsilon^{ \lambda , \lambda' , \lambda^{''}} 
\sigma^{ \lambda'} \sigma^{ \lambda^{''}} 
\:\:\:\: . 
\label{mh.7}
\eneq
\noindent
The operator $\vec{\cal T}$  
typically arises when employing the generalized JW fermionization procedure at 
a junction of three quantum spin chains \cite{crampe_trombettoni,tsvelik_Ising,gstt_1}:
it is regarded as a topological spin-1/2 operator because of its nonlocal character in both the 
chain and the site index, despite the fact that it only appears in the boundary 
Hamiltonian $H_\Delta$, which is ''concentrated'' at the common boundary (the junction) 
at $j=1$ \cite{beri_cooper,beri}. As highlighted in appendix \ref{appe_a}, 
$H_{\rm Chain}$ in Eq.(\ref{mh.4}) can be regarded as the sum of three
Kitaev Hamiltonians for a one-dimensional p-wave superconductor: on this 
analogy we will ground most of the following discussion on our system.

\section{Mapping onto the two-channel Kondo model with superconducting leads}
\label{mapping}

We  are now going to rigorously show that a junction of three off-critical quantum
Ising chains  can be mapped onto   
the  Kondo problem for a spin-1/2 impurity in contact with two superconducting 
baths (''channels''). Specifically,  we adapt to our problem the mapping procedure 
derived and discussed in Ref.\cite{tsvelik_mapping} in the case of normal 
leads. The key step is to go through the expression of the boundary Hamiltonian in terms of 
Bogoliubov   operators for a quasiparticle with energy $\epsilon$, 
$ \{ \Gamma_\epsilon \}$. This can be done by   inverting 
Eqs.(\ref{app.5}) of appendix \ref{appe_a}
and by considering that, in a spinless superconductor, one has the particle-hole
correspondence encoded in the relation $\Gamma_{ - \epsilon } = 
\Gamma_\epsilon^\dagger$, which can be explicitly 
checked from Eqs.(\ref{app.5},\ref{app.10},\ref{app.11}) of appendix \ref{appe_a}. 
Looking at the explicit  formulas for the quasiparticle wavefunctions, 
Eqs.(\ref{app.10},\ref{app.11}), one therefore obtains 

\beq
a_1^\dagger + a_1 = \sum_{ \epsilon \neq 0 } \left[ \frac{\sin ( k  + \varphi_k ) }{\sqrt{\ell + 1}} \right] 
[ \Gamma_\epsilon + \Gamma_\epsilon^\dagger ] + \left[ \frac{\sqrt{2 J^2 - 8 h^2}}{J} \right] \Gamma_{ 0 , L } 
\:\:\:\: . 
\label{map.2}
\eneq
\noindent
 $\Gamma_{ 0 , L }$ is the mode operator for the Majorana mode localized 
at the left-hand endpoint of the chain: the corresponding term in the mode 
expansion of Eq.(\ref{map.2}) only appears in the topological phase of the 
Kitaev-like Hamiltonian, corresponding to the magnetically ordered phase of 
the quantum Ising chain. As we discuss in the following, whether a term 
$\propto \Gamma_{ 0 , L }$ is present in the mode 
expansion of Eq.(\ref{map.2}), or not, does not substantially affect 
the Kondo physics of the system. Thus, in  the following of this section 
we shall just disregard it and accordingly truncate the mode expansion of 
$a_1^\dagger + a_1$ to the first term at the right-hand side of Eq.(\ref{map.2}). 
As a result, we eventually obtain 

\beq
a_1^\dagger + a_1 =  \sqrt{ \frac{2}{\ell + 1}}\: \sum_\epsilon \left[ \sin ( k + \varphi_k )  [ \Gamma_{\epsilon_k} + \Gamma_{\epsilon_k}^\dagger ] \right]
\:\:\:\: . 
\label{map.3}
\eneq
\noindent
By means of an appropriate and straightforward generalization of Eq.(\ref{map.3}), we therefore rewrite $\Sigma_1^\lambda$ in 
Eq.(\ref{mh.7}) as 

\begin{eqnarray}
\Sigma_1^\lambda &=& - \frac{ i}{  ( \ell + 1) } \: \sum_{ \lambda' , \lambda^{''}} \: \sum_{ \epsilon' , \epsilon^{''}} 
\:
\epsilon^{ \lambda , \lambda' , \lambda^{''}} \: \left[ \frac{ h^2 \sin ( k') \sin ( k^{''}) }{ \epsilon_{ k'} \epsilon_{k^{''}} } \right] 
\nonumber \\
&\times& 
[ \Gamma_{ \epsilon_{k'} , \lambda'} + \Gamma_{ \epsilon_{k'} , \lambda'}^\dagger ] [ \Gamma_{ \epsilon_{k^{''}} , \lambda^{''}} + 
\Gamma_{ \epsilon_{k^{''}} , \lambda^{''} }^\dagger ] 
\:\:\:\: .
\label{map.4}
\end{eqnarray}
\noindent
The ''bulk'' of the chains is instead described by the simple quadratic Hamiltonian given by 

\beq
H_{\rm Chain} = \sum_{\lambda = 1}^3 \: \sum_\epsilon \: \epsilon \: \Gamma_{ \epsilon , \lambda }^\dagger \Gamma_{ \epsilon , \lambda } 
\:\:\:\: . 
\label{map.5}
\eneq
\noindent
As we are now going to show, by following the main recipe presented in 
Ref.\cite{tsvelik_mapping}, it is possible to readily recover 
the total Hamiltonian $H = H_{\rm Chain } + H_\Delta$ in terms of an appropriate model 
Hamiltonian for two superconducting quasiparticle baths undergoing an appropriate Kondo-like 
interaction with the spin $\vec{\cal T}$ of  an isolated spin-1/2 impurity. To do so, let us introduce two sets of 
quasiparticle annihilation and creation operators, $\{ \gamma_{ \epsilon , a } , \gamma_{ \epsilon , a }^\dagger \}$, with 
$a = 1 , 2$, obeying the anticommutation algebra $\{ \gamma_{ \epsilon , a } , \gamma_{ \epsilon' , a'}^\dagger \} 
= \delta_{ \epsilon , \epsilon'} \delta_{ a , a'}$. Also, we choose the energy levels $\epsilon$ to cohincide with the 
eigenvalues of the single-chain Hamiltonian in Eq.(\ref{app.4}), so that the Hamiltonian for the $\gamma$-modes is given by 

\beq
H_\gamma = \sum_\epsilon \sum_a \epsilon \gamma_{ \epsilon , a}^\dagger \gamma_{ \epsilon , a  } 
\:\:\:\:.
\label{map.6}
\eneq
\noindent
Next. we   define real-space lattice 
fermion operators $\{ d_{ j , a } \}$ as 

\begin{eqnarray}
 d_{ j , 1   }&=& \sum_{ \epsilon  } \{ u_j^\epsilon \gamma_{\epsilon  , 1  } -   v_j^\epsilon  \gamma_{ \epsilon , 2}^\dagger \} \nonumber \\
 d_{j , 2 } &=& \sum_{ \epsilon  } \{  u_j^\epsilon \gamma_{\epsilon , 2}  + v_j^\epsilon  \gamma_{\epsilon , 1}^\dagger \}
 \;\;\;\; , 
 \label{map.7}
\end{eqnarray}
\noindent
with the wavefunctions $u_j^\epsilon , v_j^\epsilon$ given in Eqs.(\ref{app.10},\ref{app.11}). Now, we notice that,
going backwards to a possible lattice Hamiltonian formulation of our construction, we may construct the 
$\gamma_{ \epsilon , a }$-operators as eigenmodes of the superconducting lattice Hamiltonian $H_{\rm Eff}$, defined as 

\begin{eqnarray}
  H_{\rm Eff} =& - & \frac{J}{4} \sum_{ a = 1 , 2 } \sum_{ j = 1}^{\ell - 1} \{ d_{ j , a}^\dagger d_{ j + 1 , a } + d_{ j + 1 , a}^\dagger d_{ j , a } \} 
\nonumber \\
 &-& \frac{J}{4} \sum_{ j = 1}^{ \ell - 1 } \{ d_{ j , 1} d_{ j + 1 , 2 } - d_{ j ,2 } d_{ j + 1 , 1 } + d_{ j +1 , 2}^\dagger d_{ j , 1 }^\dagger -
 d_{ j +1 , 1}^\dagger d_{ j , 2 }^\dagger \} \nonumber \\
  &+& h \sum_{ a = 1 , 2 } \sum_{ j = 1}^\ell d_{ j , a }^\dagger d_{ j , a } 
 \:\:\:\: .
 \label{map.8}
\end{eqnarray}
\noindent
As proposed in Ref.\cite{tsvelik_mapping}, we now use the modes of $H_{\rm Eff}$ two define two independent lattice isospin operators, 
${\bf S}_j $ and ${\bf T}_j$, respectively given by 

\beq
{\bf S}_j = \frac{1}{2} \: \left[ \begin{array}{c}
                                   d_{ j , 1}^\dagger d_{ j , 2 } + d_{ j , 2}^\dagger d_{ j , 1 } \\
                                   - i (  d_{ j , 1}^\dagger d_{ j , 2 } - d_{ j , 2}^\dagger d_{ j , 1 } ) \\
d_{ j , 1}^\dagger d_{ j , 1 } - d_{ j , 2}^\dagger d_{ j , 2 }                                    
                                  \end{array} \right] 
\:\:\:\: , 
\label{map.9}
\eneq
\noindent
and by 

\beq
{\bf T}_j = \left[ \begin{array}{c}
d_{ j , 1}^\dagger d_{ j , 2}^\dagger + d_{ j , 2 } d_{ j , 1 } \\ 
- i ( d_{ j , 1}^\dagger d_{ j , 2}^\dagger - d_{ j , 2 } d_{ j , 1 } ) \\
d_{ j , 1}^\dagger d_{ j , 1} + d_{ j , 2}^\dagger d_{ j , 2} - 1 
                   \end{array} \right]
\:\:\:\: . 
\label{map.10}
\eneq
\noindent
Any component of  ${\bf S}_j $ commutes with any component of ${\bf T}_j$: therefore, the two of them  can be 
regarded as two independent spin-1/2 lattice density operators. Using them as independent channels to screen 
an isolated spin-1/2 impurity with spin $\vec{\cal T}$, coupled to the site $j=1$ by means of the 
antiferromagnetic Kondo coupling $J_K$, we may write the corresponding boundary Kondo Hamiltonian as 

\begin{eqnarray}
&& H_K = J_K \sum_{\lambda = 1}^3 \{ [ S_1^\lambda + T_1^\lambda ] {\cal T}^\lambda \}  = \nonumber \\
 & -& \frac{i J_K}{2}[  i ( d_{ 1 , 1}^\dagger - 
d_{ 1 , 1} )] [ d_{ 1 , 2}^\dagger + d_{ 1 , 2 } ] 
 {\cal T}^1 + [ d_{ 1 , 2}^\dagger + d_{ 1 , 2 } ] [ d_{ 1 , 1}^\dagger + d_{ 1 , 1 } ]  {\cal T}^2 \nonumber \\
&+&  [ d_{ 1 , 1}^\dagger  d_{ 1 , 1 } ] [  i ( d_{ 1 , 1}^\dagger - 
d_{ 1 , 1} ) ] {\cal T}^3 \}
\:\:\:\: . 
\label{map.11}
\end{eqnarray}
\noindent
From Eq.(\ref{map.11}) and from the tranformations from the $\gamma$- to the $d$-modes in Eqs.(\ref{map.7}), one 
eventually recovers the Hamiltonian $H_\Delta$, once the following identifications are performed

\begin{eqnarray}
 \Gamma_{ \epsilon , 1} + \Gamma_{ \epsilon , 1}^\dagger &\leftrightarrow&   d_{ 1 , 1}^\dagger + d_{ 1 , 1 }  \nonumber \\
  \Gamma_{ \epsilon , 2} + \Gamma_{ \epsilon , 2}^\dagger &\leftrightarrow&    i ( d_{ 1 , 1}^\dagger - 
d_{ 1 , 1} ) 
\nonumber \\ 
 \Gamma_{ \epsilon , 3} + \Gamma_{ \epsilon , 3}^\dagger &\leftrightarrow&  d_{ 1 , 2}^\dagger + d_{ 1 , 2 } 
 \:\:\:\: ,
 \label{map.12}
\end{eqnarray}
\noindent
and, of course, $J_k \leftrightarrow J_\Delta$. 
The correspondence rules in Eqs.(\ref{map.12}) complete the mapping procedure between the lattice 
two-channel superconducting-Kondo
Hamiltonian $H_{\rm Eff} + H_K$ and the model Hamiltonian for a junction of three quantum 
Ising chains. A remarkable feature of our mapping procedure is 
that   it relies on the 
construction of the spin densities for the two
independent channels as  in Eqs.(\ref{map.10},\ref{map.11}). As extensively discussed 
in Ref.\cite{tsvelik_mapping}, constructing the 
spin densities in this way implies that, if a site $j$ contains a total spin-1/2 of the 
${\bf S}_j$-operator, than it must be 
a singlet of the ${\bf T}_j$-operator, with corresponding spin equal to 0, and 
vice versa. In the ''classical'' two-channel 
Kondo problem, this is crucial a crucial point to build an effective theory for the system 
at the 2CK-fixed point which, in this regularization scheme, 
is pushed all the way down to strong coupling, such as in the 
1CK-problem \cite{tsvelik_mapping,giuliano_tagliacozzo_1}. In the following, 
we will make use of this properties to get insights of the nature of the fixed 
point toward which our system is attracted along 
the Kondo renormalization group trajectory. 

\section{Perturbative renormalizazion group analysis of the Kondo interaction}
\label{per_kondo}

\begin{figure}
\includegraphics*[width=.6\linewidth]{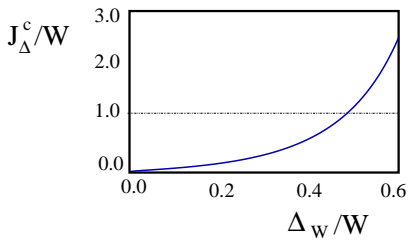}
\caption{The critical coupling $J_\Delta^c ( \Delta_w)$ as defined in Eq.(\ref{per.10}), as a function 
of $\Delta_w$ for $0 \leq \Delta_w \leq 0.6$ (in units of the bandwidth $W$). As discussed in the text, the region 
in which Kondo regime can take place corresponds to values of $J_\Delta^c / W < 1$, that is, the part of the plot 
lying below the dashed line, corresponding to $J_\Delta^c / W = 1$. } \label{critic_coupl}
\end{figure}

\begin{figure}
\includegraphics*[width=1.\linewidth]{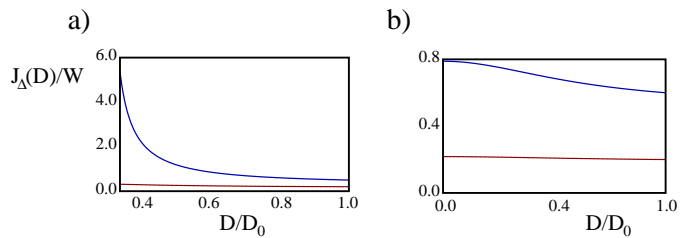}
\caption{Renormalization Group flow of $J_\Delta ( D )$ {\it vs.} $D$ for different values of 
the gap $\Delta_w$: 
{\bf a)} The curves are obtained at $\Delta_w / W = 0.3$ and for 
$J_\Delta ( D_0 ) / W = 0.6$ (blue solid curve) and $J_\Delta ( D_0 ) / W = 0.2$ (red 
dot-dashed curve). In this case $\Delta_w < \Delta_w^*$ and $J_\Delta^c (\Delta_w ) / W = 
0.5$: accordingly, $J_\Delta (D )$ flows towards strong coupling for 
$J_\Delta ( D_0 ) / W = 0.6 (> J_\Delta^c ( \Delta_w )$), while it is 
barely renormalized by the interaction when $J_\Delta ( D_0 ) / W = 0.2
(< J_\Delta^c ( \Delta_w )$; 
{\bf b)}  The curves are obtained at $\Delta_w / W = 0.6$ and for 
$J_\Delta ( D_0 ) / W = 0.6$ (blue solid curve) and $J_\Delta ( D_0 ) / W = 0.2$ (red 
dot-dashed curve). Since now $\Delta_w > \Delta_w^*$, in neither case 
$J_\Delta ( D )$ flows towards the strongly coupled regime.}  \label{flow_1}
\end{figure}

In this section, we derive the perturbative renormalization group (RG) equations  
for the running coupling $J_\Delta$. As stated above,  for the time being, we  disregard 
the zero-mode Majorana modes in the expansion of 
  $a_{1 , \lambda }^\dagger + a_{1 , \lambda}$: we will come back to 
  a discussion of their effects in the next section. To 
work out the perturbative renormalization of  $J_\Delta$, we 
 resort to the imaginary time path-integral formalism, by introducing the 
Euclidean bulk action for the chains, $S_{\rm Chain}$, given by

\beq
S_{\rm Chain} = \int  \: d \tau \: \left\{ \sum_{ \lambda = 1}^3 \sum_{ j = 1}^\ell
a_{ j , \lambda}^\dagger ( \tau ) \partial_\tau a_{ j , \lambda } ( \tau ) + 
H_{\rm Chain} ( \tau ) \right\}
\;\;\;\; , 
\label{per.1}
\eneq
\noindent
as well as the boundary action $S_\Delta$, which is given by 

\beq
S_\Delta = J_\Delta \: \int \: d \tau \: \sum_{\lambda = 1}^3 \: 
{\cal T}^\lambda ( \tau) \Sigma_1^\lambda ( \tau ) 
\:\:\:\: . 
\label{per.2}
\eneq
\noindent
Using $H_{\rm Chain}$ as noninteracting Hamiltonian, in the corresponding 
interaction representation one may present the partition function for 
the junction, ${\cal Z}$, as 

\beq
{\cal Z} = {\cal Z}_0 \langle {\bf T}_\tau \exp [ - S_\Delta ] \rangle 
\:\:\:\: , 
\label{per.3}
\eneq
\noindent
with ${\cal Z}_0 $ being the partition function for the system at $J_\Delta = 0$,
$S_\Delta$ being the boundary action in the interaction representation, and 
${\bf T}_\tau$ being the imaginary time ordering operator. Following the standard
poor man's recipe to recover the RG equation \cite{anderson,hewson_book}, we now resort 
to the frequency domain and explicitly cutoff the integration over 
frequencies at a  scale $D$, so that $S_\Delta$ can be rewritten as 

\beq
S_\Delta = \int_{ - D}^D \: \frac{  d \Omega }{ 2 \pi}  \: \sum_{ \lambda = 1}^3 \: {\cal T}^\lambda ( \Omega ) 
\Sigma_1^\lambda ( - \Omega )
\:\:\:\: , 
\label{per.4}
\eneq
\noindent
with ${\cal T}^\lambda ( \Omega )$ and $\Sigma_1^\lambda ( \Omega )$ being the 
Fourier transform of respectively ${\cal T}^\lambda ( \tau )$ and $\Sigma_1^\lambda  ( \tau)$. 
To derive the RG   equations for the running coupling, we  
 rescale the cutoff from $D $ to $ D / \kappa$, with  
$0 < \kappa - 1 \ll 1$ and, accordingly, we   split the integral in Eq.(\ref{per.4})
into   an 
integral over $ [ - D / \kappa , D / \kappa ]$ plus integrals over  values 
of $\Omega$ within  $ [ D / \kappa , D ]$ and within $ [ - D , - D / \kappa ]$. 
Leaving aside the two latter integral, as they just provide a correction to the 
total free energy. We therefore obtain, in analogy to \cite{gstt_1}

\begin{eqnarray}
S_\Delta &\to&  \int_{ - \frac{D}{\kappa}}^\frac{D}{\kappa}  \: \frac{d \Omega }{2 \pi} \: \sum_{ \lambda = 1}^3 \:  
{\cal T}^\lambda ( \Omega ) \Sigma_1^\lambda  (- \Omega ) \nonumber \\
&=& \frac{1}{\kappa}  \int_{ - D }^D  \: \frac{d \Omega }{2 \pi} \: \sum_{ \lambda = 1}^3 \:  
{\cal T}^\lambda ( \frac{ \Omega }{ \kappa } ) \Sigma_1^\lambda (- \frac{ \Omega }{ \kappa } ) 
\;\;\;\; , 
\label{per.5}
\end{eqnarray}
\noindent
which, since  $S_\Delta$   must be scale invariant, implies \cite{gstt_1}
$ \sum_{ \lambda = 1}^3 \: 
{\cal T}^\lambda ( \frac{ \Omega }{ \kappa } ) \Sigma_1^\lambda (- \frac{ \Omega }{ \kappa } ) 
= \kappa \sum_{ \lambda = 1}^3 \: {\cal T}^\lambda ( \Omega ) \Sigma_1^\lambda (- \Omega ) $.
Therefore, $J_\Delta$ takes no corrections to first order in the boundary coupling. At variance, 
to second order one finds a nonzero correction, arising from
summing over intermediate states with energies  within  
$ [ D / \kappa , D ]$ and within $ [ - D , - D / \kappa ]$. Performing the integration, 
one eventually obtains  that, to leading 
order in $J_\Delta$ (corresponding to one-loop order in the expansion of the action 
in $S_\Delta$), $S_\Delta$ is corrected according to $S_\Delta \to S_\Delta 
+\delta S_\Delta^{(2)}$, with  \cite{gstt_1}

\beq
\delta S_\Delta^{(2)}
 =   J_\Delta^2
 \int_{ - D}^D \: \sum_{ \lambda = 1}^3 \: 
 \frac{d \Omega }{2 \pi} \: {\cal T}^\lambda ( \Omega ) \Sigma_1^\lambda  
(- \Omega )  [ \Gamma  ( D ) +    \Gamma  ( -  D  ) ] 
D ( 1 - \kappa^{-1} ) 
\:\:\:\: .
\label{per.6}
\eneq
\noindent
The function $\Gamma ( \Omega )$ in Eq.(\ref{per.6}) is defined to be 
  the Fourier-Matsubara transform of  
$\Gamma  ( \tau ) = G ( \tau ) g ( \tau )$, with $ g ( \tau ) = {\rm sgn} 
( \tau )$ being the $\sigma$-fermion 
Green's function $g ( \tau ) = \langle {\bf T}_\tau [ \sigma ( \tau ) 
\sigma ( 0 ) ] \rangle$ and $G  ( \tau )$ being the imaginary time ordered
Green's function (effectively independent of $\lambda$, due to the equivalence 
between the three chains)

\beq
G ( \tau ) = - \langle {\bf T}_\tau \{ [ a_{ 1 , \lambda}^\dagger ( \tau ) + 
a_{ 1 , \lambda } ( \tau ) ] [ a_{ 1 , \lambda}^\dagger ( 0 ) + 
a_{ 1 , \lambda } ( 0 ) ]  \} \rangle 
\:\:\:\: .
\label{per.7}
\eneq
\noindent
From Eqs.(\ref{per.6},\ref{per.7}) one may eventually derive the RG equations for 
the running coupling $J_\Delta ( D )$ in the form 

\beq
\frac{ d J_\Delta ( D ) }{d \ln \left( \frac{D_0}{D} \right) } = J_\Delta^2 ( D ) \rho ( D)
\:\:\:\: , 
\label{per.8}
\eneq
\noindent
with, in the specific system we are focusing on, $\rho ( D )$ being given by 

\beq
\rho ( D ) = \frac{16}{\pi J^2} \: \int_{ \Delta_w}^{W} \: d \epsilon \: \left[ \frac{  D^2}{( \epsilon^2 + D^2 )^2 }
\right] \sqrt{( \epsilon^2 - \Delta_w^2 ) ( W^2 - \epsilon^2 ) }
\:\:\:\: . 
\label{per.9}
\eneq
\noindent
In Eq.(\ref{per.9}), we used $\Delta_w$ to denote the single-fermion excitation gap, as discussed in appendix 
\ref{appe_a}, while $ W = \frac{J}{2} + | h | $ is the energy at the band edge in the single-fermion spectrum and 
$D_0$ is a ${\cal O} ( W )$ high-energy reference cutoff. 
On integrating Eq.(\ref{per.9}), one may therefore infer whether the system crosses over towards the Kondo regime, 
despite the presence of a nonzero gap $\Delta_w$ in the spectrum, and, if that is the case, what is the corresponding
(Kondo) temperature scale at which the crossover takes place. To analyze the onset of the Kondo regime, we follow
the technique highlighted in \cite{luca_dellanna}. Specifically, we introduce the ($\Delta_w$-dependent) ''critical coupling'' $J_\Delta^c ( \Delta_w ) $, 
defined as 

\beq
J_\Delta^c (\Delta_w ) = \Biggl\{ \int_{0}^{D_0} \: \rho ( x ) \frac{d x}{x} \Biggr\}^{-1}
\:\:\:\: . 
\label{per.10}
\eneq
\noindent
(Note that, in the definition of $J_\Delta^c$, we stressed the dependence on the gap $\Delta_w$. This is 
a basic feature of our ''off-critical'' model, which makes the main difference between the case we investigate
here and the critical limit, extensively discussed in \cite{tsvelik_Ising,gstt_1}.)
Having introduced the critical coupling, the solution to Eq.(\ref{per.8}) can be rewritten as 

\beq
J_\Delta ( D ) = \frac{J_\Delta ( D_0 ) J_\Delta^c ( \Delta_w ) }{ J_\Delta ( D_0 ) - J_\Delta^c ( \Delta_w ) +
J_\Delta ( D_0 ) J_\Delta^c ( \Delta_w ) \: \int_0^D \rho ( x ) \frac{d x}{x}}
\:\:\:\: . 
\label{per.11}
\eneq
\noindent
Within standard poor man's approach to Kondo problem, the onset of the Kondo regime is signaled by 
the appearance of a scale $D_K$ at which $J_\Delta ( D )$ diverges. At $D = D_K$, therefore, the denominator 
of the expression at the right-hand side of Eq.(\ref{per.11}) must be equal to 0, which is only possible 
if $J_\Delta ( D_0 ) \geq J_\Delta^c ( \Delta_w )$. 
This observation implies that Kondo effect does definitely not take place 
whenever $J_\Delta^c ( \Delta_w ) / W > 1$. At variance, for  $J_\Delta^c ( \Delta_w ) / W < 1$ the 
crossover to Kondo regime can take place within an appropriate range of values of $J_\Delta ( D_0 )$, 
provided the Kondo crossover scale $D_K$, though substantially 
lower than $W$, is still $ > \Delta_w$, so to have a nonzero fermion density screening the isolated magnetic
impurity at the scale $D_K$ \cite{avishai,choi_1,campagnano_1,choi_2}. In Fig.\ref{critic_coupl}, we 
plot $J_\Delta^c ( \Delta_w ) / W$ as a function of $\Delta_w$.  The dashed horizontal line marks the set of points
corresponding to $J_\Delta^c / W = 1$. Consistently with what we discuss before, we expect that Kondo regime is 
fully suppressed by the gap in the single-fermion spectrum throughout all the region with $J_\Delta^c / W > 1$, that is, 
for $\Delta_w > \Delta_w^* \approx 0.5 W$.
\noindent
To check the consistency between the RG flow of the running coupling 
 and Eq.(\ref{per.11}), we numerically compute $J_\Delta ( D )$ {\it vs.} $D$ by 
 integrating Eq.(\ref{per.8}) for different values of $\Delta_w$. We
 report the corresponding curves in Fig.\ref{flow_1}a): specifically, 
 we find that, for $\Delta_w = 0.3$ (that is, much lower than $\Delta_w^*$), 
 $J_\Delta ( D)$ either flows towards the strongly coupled regime, or not, according 
 to whether $J_\Delta ( D_0 ) > J_\Delta^c ( \Delta_w ) \: (\approx 0.51)$, or 
$J_\Delta ( D_0 ) < J_\Delta^c ( \Delta_w )$.  At variance,  as it
clearly appears in Fig.\ref{flow_1}b), for $\Delta_w > \Delta_w^*$, 
$J_\Delta ( D )$ is barely renormalized by the Kondo interaction and shows no 
evidence of nonperturbative flow towards strong coupling.
\noindent
Once the conditions under which the onset of the Kondo regime take  place, it 
  becomes  important to infer the dependence of the corresponding Kondo temperature
scale $T_K$ on both $J_\Delta ( D_0 )$ and $\Delta_w$. By definition, one sets 
$T_K = D_K / k_B$, where $k_B$ is the Boltzmann constant and $D_K$ is the scale at
which the denominator of Eq.(\ref{per.11}) becomes equal to 0. Then,  $D_K$ is formally 
given by the equation 

\begin{eqnarray}
&& 1 + J_\Delta ( D_0 ) \left[ \frac{32}{\pi \left( 1 + \frac{\Delta_w^2}{W^2} \right)} \right] \times 
\label{per.12} \\
&& \int_{ \frac{\Delta_w }{W}}^1 \: d u \: \sqrt{u^2 - \frac{\Delta_w^2}{W^2}} \sqrt{1 - u^2}
\left[ \frac{1}{u^2 + 1 } - \frac{1 }{u^2 + \frac{D_K^2}{W^2}} \right] = 0 \nonumber
\:\:\:\: . 
\end{eqnarray}
\noindent
As stated before, in order for the   Kondo regime to take place, it is 
important that the condition $D_K / \Delta_W \gg 1$ is satisfied. Within such an 
hypothesis, we can therefore simply approximate the factor $\sqrt{u^2  - \frac{\Delta_w^2}{W^2}} $
in Eq.(\ref{per.12}) with $u$. In addition, as Kondo physics is mostly a low-energy effect, 
we may also approximate $\sqrt{1 - u^2}$ simply with 1. As a result, we eventually obtain 

\beq
T_K [ J_\Delta ( D_0 ) ; \Delta_w ] \approx D_0 \exp \left[ -  
\frac{\pi \left( W^2 + \Delta_w^2 \right)}{32 W J_\Delta ( D_0 ) } \right]
\:\:\:\: , 
\label{per.13}
\eneq
\noindent
with the cutoff $D_0 \sim W$. As $\Delta_w \to 0$, Eq.(\ref{per.13}) gives back the 
result for the Ising limit, once the proper correspondence between the various parameters
has been traced out \cite{gstt_1}. As s general result, Eq.(\ref{per.13}) encodes a remarkable
''Kondo temperature renormalizazion'', namely,    on increasing $\Delta_w$, 
one sees a remarkable reduction of $T_K  $, which is consistent with 
the expected competition between Kondo physics and gapped spectrum \cite{avishai,choi_1,campagnano_1,choi_2}.
 Specifically, Eq.(\ref{per.13}) is expected to apply to the regime in which a nonzero $\Delta_w$ does not suppress Kondo
effect, that is, for $D_K / \Delta_w \gg 1$. The suppression of Kondo effect with increasing 
$\Delta_w$ can instead be numerically derived, by using the integrated RG flow in 
Eq.(\ref{per.11}) to estimate $T_K $ {\it vs.} $\Delta_w$ at fixed $J_\Delta ( D_0 )$ and 
$T_K $ {\it vs.} $J_\Delta ( D_0 )$ at fixed $\Delta_w$. As a result, one obtains plots such as the 
ones we show in Fig.\ref{ktemp} at the system parameters chosen as discussed in the caption. In particular, 
the suppression of Kondo effect either on increasing $\Delta_w$ at fixed $J_\Delta ( D_0 ) $, or on 
decreasing $J_\Delta ( D_0 )$ at fixed $\Delta_w$ is evidenced by
the fact that the curve $T_K [ J_\Delta ( D_0 ) ; \Delta_w ]$ becomes 
constantly zero above (below) a 
critical value of $\Delta_w \; (J_\Delta ( D_0 )) $ at fixed $J_\Delta ( D_0 ) \; (\Delta_w ) $.

\begin{figure}
\includegraphics*[width=1.\linewidth]{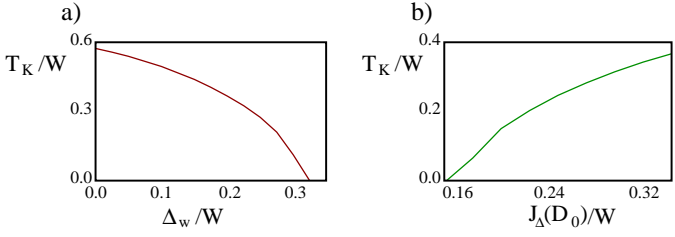}
\caption{Full plot of the Kondo temperature as a function of  $\Delta_w$ at fixed 
$J_\Delta ( D_0 )$, and  
 of $J_\Delta (D_0 )$ at fixed $\Delta_w$: 
{\bf a) } $T_K / W $ {\it vs.} $\Delta_w / W$ at $J_\Delta ( D_0 ) / W = 0.35 $ ; 
{\bf b) } $T_K / W $ {\it vs.} $J_\Delta ( D_0 ) / W$ at $ \Delta_w / W = 0.2 $.  }  \label{ktemp}
\end{figure}
\noindent

\section{Majorana modes and onset of the Kondo regime}
\label{majo_kondo}

In this section we extend the perturbative RG analysis to additional terms in 
$H_\Delta$ which arise due to  emergence of Majorana modes (MM)s   at
the endpoints of the chain. In particular, we  show that, under the required assumptions for 
recovering Kondo effect, these terms do not affect  the onset of Kondo physics.  
As a starting point, we note that, on including the zero-mode MM in the mode expansion of 
$a_{1 , \lambda} + a_{ 1 , \lambda }^\dagger$,   Eq.(\ref{map.4}) 
for $\Sigma_1^\lambda$ is modified to

\beq
\Sigma_1^\lambda = \rho^2 {\cal R}^\lambda + \rho \omega_1^\lambda  + \bar{\Sigma}_1^\lambda
\:\:\:\: , 
\label{mmo.1}
\eneq
\noindent
with $\bar{\Sigma}_1^\lambda$ contributed by nonzero modes and given by the mode expansion in 
Eq.(\ref{map.4}), $\rho = \frac{\sqrt{2 J^2 - 8 h^2}}{J} = 2 \sqrt{2}
\frac{\sqrt{\Delta_w W}}{J}$, and

\begin{eqnarray}
 {\cal R}^\lambda &=& - \frac{i}{2} \sum_{ \lambda' , \lambda^{''}} \: \epsilon^{ \lambda , \lambda' , \lambda^{''}} 
 \: \Gamma_{ 0 , L}^{\lambda'} \Gamma_{ 0 , L}^{ \lambda^{''}} \nonumber \\
 \omega_1^\lambda &=& - i \sum_{ \lambda' , \lambda^{''}} \: \epsilon^{ \lambda , \lambda' , \lambda^{''}} \Gamma_{ 0 , L}^{\lambda^{''}} \times \nonumber \\
 && \left[  \sum_{ \epsilon }  \left( \frac{ h \sin ( k )  }{ \epsilon_{ k }  } \right) 
[ \Gamma_{ \epsilon_{k } , \lambda'} + \Gamma_{ \epsilon_{k  , \lambda'}}^\dagger ]  \right] 
\:\:\:\: . 
\label{mmo.2}
\end{eqnarray}
\noindent
On inserting Eq.(\ref{mmo.1}) into the expression of $H_\Delta$, one eventually finds 

\beq
H_\Delta = H_\Delta^{(0)} + H_\Delta^{(1)} + \bar{H}_\Delta
\:\:\:\: , 
\label{mmo.3}
\eneq
\noindent
with 

\begin{eqnarray}
 H_\Delta^{(0)} &=&  J_0 \: \sum_{ \lambda = 1}^3 {\cal T}^\lambda {\cal R}^\lambda \nonumber \\
 H_\Delta^{(1)} &=& J_1 \: \sum_{ \lambda = 1}^3 {\cal T}^\lambda \omega_1^\lambda  \nonumber \\
 \bar{H}_\Delta &=& \bar{J} \: \sum_{ \lambda = 1}^3 {\cal T}^\lambda \bar{\Sigma}_1^\lambda 
 \:\:\:\: .
 \label{mmo.4}
\end{eqnarray}
\noindent
$J_0 , J_1 $ and $\bar{J}$ in Eq.(\ref{mmo.4}) are three in principle independent running 
couplings which, at the bare level, are respectively given  by 
$J_0 = \rho^2 J_\Delta$, $J_1 =  \rho J_\Delta $, and $\bar{J} = J_\Delta$. 
$\bar{H}_\Delta$ in Eq.(\ref{mmo.4}) is basically the same operator as one gets  in the 
absence of MMs.  We have performed the 
full perturbative RG analysis of the corresponding running coupling strength $\bar{J} ( D )$ in the previous 
section and have concluded that, under appropriate conditions on  $\Delta_w$   
 $J ( D_0 )$, the system can develop Kondo effect, corresponding to a marginally relevant rise of 
$\bar{J} ( D)$, as $D $ is lowered from $D_0$ towards $D_K$. $H_\Delta^{(0)}$ takes the form of a 
''RKKY''-like coupling between two topological spin-1/2 operators, $\vec{\cal T}$ determined by 
the Klein factors $\sigma^\lambda$, and  $\vec{\cal R}$, determined by 
 the MMs $\Gamma_{0 , L}^\lambda$ as from Eq.(\ref{mmo.2}). Based on dimensional counting 
arguments for boundary interaction terms \cite{cardy_book}, one expects that, on lowering $D$, the corresponding 
running coupling $J_0 ( D )$ scales as $J_0 ( D ) = J_0 ( D_0 ) \frac{D_0}{D}$ and, similarly, that
  $J_1 ( D )$  scales as 
$J_1 ( D ) = J_1 ( D_0 ) \left( \frac{D_0}{D} \right)^\frac{1}{2}$.
Apparently, on lowering $D$, this implies a rise of both $J_0 ( D )$ and $J_1 ( D )$ 
faster than $\bar{J } ( D )$. However, one has to recall that, 
by definition, the scaling must be terminated at the scale $D = D_K$. At such a scale, 
  one obtains 

\beq
J_0 ( D_K ) = \rho^2 J_\Delta \frac{D_0}{D_K} \sim \frac{W^2}{J^2} \frac{\Delta_w}{D_K} \: J_\Delta
\:\:\:\: . 
\label{mmo.5}
\eneq
\noindent
Within the magnetically ordered phase, the condition $ | 2 h | < J$ implies $ | W | / J \leq 1$. Moreover, 
our assumtion on the onset of the Kondo regime implies $ \Delta_w / D_K \ll 1$, which eventually 
yields $ J_0 ( D_K ) \ll J_\Delta$. By means of a similar argument, one readily concludes that 
$J_1 ( D_K ) \ll J_\Delta$, as well. As a result, all the way down to $D = D_K$, $H_\Delta^{(0)}$ and $H_\Delta^{(1)}$ 
merely provide a perturbative, small additional boundary interaction, which we neglect, against the 
relevant Kondo-like interaction $\bar{H}_\Delta$. It would be interesting to analyize whether it 
is possible to modify the Hamiltonian $H_\Delta$ so to eventually make the RKKY-interaction to be 
relevant, in the magnetically ordered phase. In fact, this would provide a tool to monitor the emergence of 
MMs in terms of pertinent modifications in the boundary phase diagram associated to $H_\Delta$
(suppression of Kondo effect). This, however, 
lies outside the scope of this work, and we plan to discuss it in a future publication. 
We thus conclude that,  at least down to 
the scale $D = D_K$, the MMs do not provide 
sensible modification to the Kondo RG flow of the boundary coupling $J_\Delta 
( D )$, which makes the 
discussion of the previous section to be equally valid for the paramagnetic, 
as well as for the ferromagnetic
phase of the spin chains. 

\section{Description of the strongly coupled Kondo fixed point}
\label{strocou}

\begin{figure}
\includegraphics*[width=.6\linewidth]{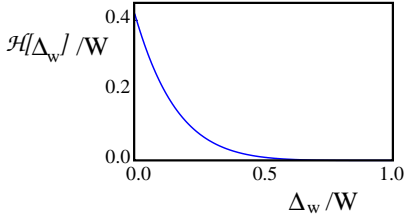}
\caption{ Plot of the function ${\cal H} [ \Delta_w ]$ {\it vs.} $\Delta_w$ (both quantities are
measured in units of $W$).}  \label{hdelta}
\end{figure}

In the previous sections we have shown that Kondo effect in our system is 
recovered whenever, at a given value of $\Delta_w$, one 
has $J_\Delta ( D_0 ) > J_\Delta^c ( \Delta_w )$, implying a flow of 
the boundary interaction towards the strongly coupled  Kondo fixed point (KFP).  
In this section, we provide a  description of the sytem at the KFP. To do so,  we  
combine the formal description of the KFP in the gapless case 
developed in \cite{tsvelik_mapping,giuliano_tagliacozzo_1} with a pertinently 
modified version of the projection 
variational approach used to study the single-channel KFP with 
superconducting leads  \cite{campagnano_1}.
The starting point is the observation that the topological spin $\vec{\cal T}$ is 
only coupled to the total spin density at the site $j=1$. 
As a consequence of the properties of the two spin-1/2 operators ${\bf S}_j$ and 
${\bf T}_j$ introduced in section \ref{mapping},    $j=1$ hosts  total spin-1/2  of ${\bf S}_1$
and total spin-0 of ${\bf T}_1$, or 
vice versa. As a result, when coupled to both spins by means of the 2CK-like interaction in 
Eq.(\ref{map.11}), $\vec{\cal T}$  can give rise to either a total 0-spin spin
singlet, or to a total spin-1 spin triplet state. In the noninteracting limit $J_\Delta = 0$, the actual groundstate of 
the system is an equally-weighted mixture of singlet- and triplet-states. As $J_\Delta$ is 
turned on, we expect that, the larger is $J_\Delta$, the higher is the relative weight of the 
singlet states versus the triplet states \cite{campagnano_1}. To formally ground
this observation, we define the operator ${\cal P}_g = 
{\bf I} + g \sum_{ \lambda = 1}^3 {\cal T}^\lambda 
\Sigma_1^\lambda $. For $g= - 4$, ${\cal P}_g$ fully projects out the localized triplet. 
To set the ''optimal'' value of $g$ at a given $J_\Delta$, one 
employs a variational procedure, consisting in evaluating the average value of the 
total Hamiltonian onto the projected out state at fixed $g$,
${\cal E} [ J_\Delta ; \Delta_w ;  g ]$,
and, at a given $J_\Delta$, in choosing $g$ so to minimize ${\cal E} [ J_\Delta  ; \Delta_w ; g ]$.
This determines a curve $g ( J_\Delta )$, from which one can infer what is the 
optimal state as $J_\Delta \to \infty$ (KFP).  We define the 
projected state $ | \Psi \rangle_g$ as 

\beq
| \Psi \rangle_g = \frac{ {\cal P}_g | {\rm GS} ; \Uparrow \rangle }{ \sqrt{ \langle {\rm GS} ; 
\Uparrow | {\cal P}_g^2 | {\rm GS } ; \Uparrow \rangle }}
\:\:\:\: , 
\label{proj.1}
\eneq
\noindent
with $ | {\rm GS} ; \Uparrow \rangle = | {\rm GS} \rangle \otimes | \Uparrow \rangle$ and
$ | {\rm GS} \rangle$ being the groundstate of the chain Hamiltonian in Eq.(\ref{map.5}),
while $ | \Uparrow \rangle$ being one of the two eigestates of $ {\cal T}^z $ (we expect our
final result not to sensibly depend on the choice of the initial state to project out, which 
enables us to  arbitrarely choose the initial state). 
It is simple, now, to prove that one gets 

\beq
\langle {\rm GS} ; 
\Uparrow | {\cal P}_g^2 | {\rm GS } ; \Uparrow \rangle = 1 + \frac{3 g^2}{16}
\:\:\:\: . 
\label{proj.2}
\eneq
\noindent
Moreover,   one also obtains 

\begin{eqnarray}
&& \langle {\rm GS} ; 
\Uparrow | {\cal P}_g H_{\rm Chains} {\cal P}_g  | {\rm GS } ; \Uparrow \rangle  \nonumber \\
&=& E_{\rm GS} \left\{ 1 + \frac{3 g^2}{16} \right\} + 3 g^2 \Psi_1 [ \Delta_w ] \Psi_2 [ \Delta_w ] 
\:\:\:\: , 
\label{proj.3}
\end{eqnarray}
\noindent
with 

\begin{eqnarray}
 \Psi_1 [ \Delta_w ] &=& \left[ \frac{1}{\pi ( 1 + \Delta_w )^2 } \right] \: 
 \int_{ \Delta_w}^W \: \frac{d \epsilon}{\epsilon} \: \sqrt{\epsilon^2 - \Delta_w^2}
 \sqrt{W^2 - \epsilon^2} \nonumber \\
  \Psi_2 [ \Delta_w ] &=& \left[ \frac{1}{\pi ( 1 + \Delta_w )^2 } \right] \: 
 \int_{ \Delta_w}^W \:  d \epsilon  \: \sqrt{\epsilon^2 - \Delta_w^2}
 \sqrt{W^2 - \epsilon^2} \:\: , \nonumber \\
 \:\:
 \label{proj.4}
 \end{eqnarray}
 \noindent
 and $E_{\rm GS} = \langle {\rm GS} | H_{\rm Chain} | {\rm GS} \rangle$.  
In order to find out the last contribution to the averaged energy, 
we need the following identities

\begin{eqnarray}
 [ \sum_{ \lambda = 1}^3 \Sigma_1^\lambda {\cal R}^\lambda ]^2 &=& \frac{3}{16} - 
 \frac{1}{4} \sum_{ \lambda = 1}^3 \Sigma_1^\lambda {\cal T}^\lambda  \nonumber \\
  [ \sum_{ \lambda = 1}^3 \Sigma_1^\lambda {\cal T}^\lambda ]^3 &=& 
  - \frac{3}{64} + \frac{1}{4} \sum_{ \lambda = 1}^3 \Sigma_1^\lambda {\cal T}^\lambda
  \:\:\:\: . 
  \label{proj.5}
\end{eqnarray}
\noindent
Therefore,  we obtain 

\beq
\langle  {\rm GS} ; \Uparrow | {\cal P}_g H_\Delta {\cal P}_g | {\rm GS} ; \Uparrow \rangle
 = \left[  \frac{3 g}{8} - \frac{3 g^2}{64} \right] J_\Delta
 \:\:\:\: , 
 \label{proj.6}
 \eneq
 \noindent
so that we  eventually get 

\beq
{\cal E} [ J_\Delta ; \Delta_w ; g ] = E_{\rm GS} + \frac{3 g^2 \Psi_1 [ \Delta_w ] \Psi_2 [ \Delta_w ] + 
\frac{3}{8} g \left( 1 - \frac{g}{4} \right) J_\Delta}{1 + \frac{3 g^2}{16}} 
\:\:\:\: .
\label{proj.7}
\eneq
\noindent
The condition $\partial_g {\cal E} [ J_\Delta ; \Delta_w ; g ] = 0 $ is satisfied by 
either setting $g = g_1 [ J_\Delta ; \Delta_w ] $, or $g=g_2 [ J_\Delta ; \Delta_w ] $,
with 

\begin{eqnarray}
 g_1 [ J_\Delta ; \Delta_w ] &=&  \frac{ 2 {\cal H} [ \Delta_w  ] - J_\Delta - 2 \sqrt{{\cal H}^2 [ \Delta_w ] - 
 {\cal H} [ \Delta_w ] J_\Delta + J_\Delta^2 }}{ 3 J_\Delta / 4 } \nonumber \\
  g_2 [ J_\Delta ; \Delta_w ] &=&  \frac{ 2 {\cal H} [ \Delta_w  ] - J_\Delta + 2 \sqrt{{\cal H}^2 [ \Delta_w ] - 
 {\cal H} [ \Delta_w ] J_\Delta + J_\Delta^2 }}{ 3 J_\Delta / 4 } \:\: ,\nonumber \\
 \:\:
 \label{proj.8}
\end{eqnarray}
\noindent
and  ${\cal H} [ \Delta_w ] = 16 \Psi_1 [ \Delta_w ] \Psi_2 [ \Delta_w ] $.
In Fig.\ref{hdelta}, we plot ${\cal H} [ \Delta_w ] / W $ {\it vs.} $\Delta_w / W$. We see that
${\cal H} [ \Delta_w ]$ keeps finite at not-too-large values of $\Delta_w$. Therefore, 
we may readily compute $g_j^*  = \lim_{J_\Delta \to \infty }  g_j [ J_\Delta ; \Delta_w ] $
from Eqs.(\ref{proj.8}), obtaining $g_1^* = - 4 , g_2^* = \frac{4}{3}$. The latter value corresponds 
to projecting out the singlet and, therefore, it maximizes ${\cal E}$. Therefore, we take for good
the former value which, as expected, corresponds to fully projecting out the triplet and 
to having a localized singlet at the effective magnetic impurity. We therefore conclude that having 
a nonzero $\Delta_w$ does not spoil  Nozi\`ere's   picture of the system's groundstate as a localized 
spin singlet at the impurity. As a result, we may readily describe the system's groundstate as a twofold degenerate
singlet, formed by $\vec{\cal T}$ and either one between ${\bf S}_1$ and ${\bf T}_1$ which can be simply described
within our approach as discussed in \cite{tsvelik_mapping,giuliano_tagliacozzo_1}.

\noindent

\section{Discussion and conclusions}
\label{conclusions}
 
In this paper we rigorously prove that a junction 
of three off-critical quantum Ising chains can be  regarded as a quantum spin chain realization 
of the two-channel spin-1/2 overscreened Kondo 
effect with two superconducting leads. By making a combined use of 
a pertinently adapted version of poor man's perturbative RG approach 
to the Kondo problem \cite{anderson,hewson_book} and of the   variational approach to the strong coupling fixed
point based on progressively projecting out from the Hilbert space states different from a localized 
singlet at the impurity site \cite{campagnano_1,giuliano_tagliacozzo_1}, we show that,   on lowering the 
reference energy scale $D$, the system flows all the way down to 2CK-fixed point.  

Our result paves the way to the possibility of realizing and studying in a controlled 
setting   2CK-effect with 
superconducting lead, so far never considered in a solid-state quantum dot device.
In fact, our proposed device appears to be within the reach of nowadays technology 
in nanostructures and could be engineered by means, for instance, of a pertinent 
Josephson junction network \cite{giuliano_sodano_epl}. 
To detect the effect in the quantum spin chain system, one may  look, for instance, 
at  the scaling with $D$ of the local magnetization at the endpoints
of the chains (such as discussed in \cite{gstt_1}). Alternatively, 
in the Josephson junction  network realization of the system one can in
principle detect the effect  by means of an appropriate dc Josephson current
measurement \cite{giuliano_sodano_epl}. 

Besides the theoretical interest, our results are potentially of interest for what concerns quantum 
entanglement properties of the system, which suggests a possible further development of our research 
towards quantum computation related issues. Finally, it would be interesting to study how the effect 
is modified by e.g. the introduction of disorded in the chains, by inhomogeneities in the boundary 
couplings, etc. Such a topics, though interesting, lie nevertheless beyond the scope of this work
and we will possibly reserve them for a further publication.

 \vspace{0.5cm}

We thank  P. Sodano and A. Trombettoni for  insightful discussions during 
the preparation of this work. 
 
\appendix 

\section{Fermionization and explicit solution of a single quantum Ising chain 
with open boundary conditions}
\label{appe_a}

 In this appendix, we review the Jordan-Wigner fermionization procedure applied to 
a single QIC with open boundary conditions and eventually present the exact solution of
the model Hamiltonian in terms of Jordan-Wigner fermions. The Hamiltonian for a single 
chain is given by 

\beq
H_1 = - J \sum_{ j = 1}^{\ell - 1 }S_{ j + 1 }^x S_j^x  + h \sum_{ j = 1}^\ell 
S_j^z 
\:\:\:\: . 
\label{app.1}
\eneq
\noindent
The Jordan-Wigner tranformation \cite{jordan_wigner}
allows us for trading the bosonic Hamiltonian $H_1$
for a fully fermionic one, by introducing a set of 
spinless lattice fermions $\{ a_j , a_j^\dagger \}$, 
obeying the basic anticommutation relations $\{ a_j , a_{j'}^\dagger \} = 
\delta_{ j , j'}$. The relations between the bosonic spins and the fermionic operators 
are determined so to preserve the correct (anti)commutation relations. They are 
therefore given by 

\begin{eqnarray}
S_j^+ &=& a_j^\dagger \: e^{ i \pi \sum_{ r = 1}^{ j - 1 } a_r^\dagger a_r } \nonumber \\
S_j^- &=& a_j  \: e^{ i \pi \sum_{ r = 1}^{ j - 1 } a_r^\dagger a_r } \nonumber \\
S_j^z &=& a_j^\dagger a_j - \frac{1}{2}
\:\:\:\: , 
\label{app.2}
\end{eqnarray}
\noindent
with 

\begin{eqnarray}
 S_j^x &=& \frac{1}{2} [ S_j^+ + S_j^- ] \nonumber \\
  S_j^y &=& \frac{-i}{2} [ S_j^+ - S_j^- ]
  \:\:\:\: . 
  \label{app.3}
\end{eqnarray}
\noindent
On inserting  Eqs.(\ref{app.2}) into the Hamiltonian in Eq.(\ref{app.1}), 
we readily resort to the fully fermionized version of $H_1$, given by 

\begin{eqnarray}
&& H_1 = - \frac{J}{4} \: \sum_{ j = 1}^{\ell - 1 } \{ a_j^\dagger a_{ j + 1 } + 
a_{ j + 1 }^\dagger a_j \} \nonumber \\
&-& \frac{J}{4} \: \sum_{ j = 1}^{\ell - 1 } \{ a_j  a_{ j + 1 } + 
a_{ j + 1 }^\dagger a_j^\dagger \} + h \sum_{ j = 1}^\ell
a_j^\dagger a_j 
\:\:\:\: . 
\label{app.4}
\end{eqnarray}
\noindent
The Hamiltonian in Eq.(\ref{app.4}) is
Kitaev's model Hamiltonian for a one-dimensional p-wave superconductor
\cite{kitaev}, with the various parameter (in the notation of 
\cite{kitaev}) chosen as $w = \Delta = \frac{J}{4}$, $\mu = - h$. To 
explicitly determine the energy eigenmodes of $H_1$, $\Gamma_\epsilon$, we 
assume that they take the form 

\begin{eqnarray}
 \Gamma_\epsilon &=& \sum_{ j = 1}^\ell \{ [ u_j^\epsilon ]^* a_j + [ v_j^\epsilon ]^*
 a_j^\dagger \} \nonumber \\
  \Gamma_\epsilon^\dagger &=& \sum_{ j = 1}^\ell \{  v_j^\epsilon a_j + u_j^\epsilon 
 a_j^\dagger \}
 \:\:\:\: , 
 \label{app.5}
\end{eqnarray}
\noindent
with $u_j^\epsilon , v_j^\epsilon$ being the lattice version of the 
quasiparticle wavefunction solving the Bogoliubov-de Gennes (BDG) equations 
for a superconductor \cite{degennes_book}. Imposing the commutation 
relation $ [ \Gamma_\epsilon , H_1 ] = \epsilon \Gamma_\epsilon$, one 
therefore obtains the BDG equations for $ ( u_j^\epsilon , v_j^\epsilon )$, in
the form 

\begin{eqnarray}
 \epsilon u_j^\epsilon &=& - \frac{J}{4} \{ u_{ j + 1 }^\epsilon + u_{ j - 1 }^\epsilon \} 
 + \frac{J}{4} \{ v_{ j + 1 }^\epsilon - v_{j-1}^\epsilon \}  +  h u_j^\epsilon \nonumber \\
  \epsilon v_j^\epsilon &=&   \frac{J}{4} \{ v_{ j + 1 }^\epsilon + v_{ j - 1 }^\epsilon \} 
- \frac{J}{4} \{ u_{ j + 1 }^\epsilon - u_{j-1}^\epsilon \} -  h v_j^\epsilon
\:\:\:\: , 
\label{app.6}
\end{eqnarray}
\noindent
for $1 < j < \ell$, supplemented with the boundary conditions at $j=1 , \ell$ given by 
(see \cite{gstt_1} for a detailed discussion of the implementation of open boundary conditions
within the fermionic description of open quantum spin chains) 

\beq
u_0^\epsilon +v_0^\epsilon = u_{\ell + 1}^\epsilon - v_{\ell + 1}^\epsilon = 0 
\:\:\:\: . 
\label{app.7}
\eneq
\noindent
In solving Eqs.(\ref{app.6}) in combination with the boundary conditions in 
Eq.(\ref{app.7}), we  
look for solutions   of the form 

\beq
\left[ \begin{array}{c}
u_j^\epsilon \\ v_j^\epsilon          
       \end{array} \right] = \left[ \begin{array}{c}
                                     u_k \\ v_k 
                                    \end{array} \right] 
                                    e^{ i k j } 
\:\:\:\: . 
\label{app.9}
\eneq
\noindent
At a given $k$, we then find two independent solutions at energy 
$\pm \epsilon_k$, with $\epsilon_k = \sqrt{\frac{J^2}{4} + h^2 -  J h \cos ( k )}$, and 
the two solutions respectively given by 

\beq
\left[ \begin{array}{c}
        u_j^\epsilon \\ v_j^\epsilon 
       \end{array} \right]_+ = \sqrt{ \frac{2}{\ell + 1}} \: 
       \left[ \begin{array}{c}
                                       \cos \left( \frac{\varphi_k}{2} \right)  
                                       \sin \left[ k j + \frac{\varphi_k}{2} \right] \\
-  \sin \left( \frac{\varphi_k}{2} \right)  
                                       \cos \left[ k j + \frac{\varphi_k}{2} \right]                                        
                                      \end{array} \right]  
\;\;\;\; , 
\label{app.10}
\eneq
\noindent
for the positive energy solution, with the allowed values of $k$ determined by the 
secular equation

\beq
\sin \left[ k ( \ell + 1 ) + \varphi_k \right] = 0 
\:\:\:\: , 
\label{app.10bis}
\eneq
\noindent
and 

\beq
\left[ \begin{array}{c}
        u_j^\epsilon \\ v_j^\epsilon 
       \end{array} \right]_- = \sqrt{ \frac{2}{\ell + 1}} \: 
       \left[ \begin{array}{c}
 \sin \left( \frac{\varphi_k}{2} \right)  
                                       \cos \left[ k j + \frac{\varphi_k}{2} \right]  \\  
  
                                     -   \cos \left( \frac{\varphi_k}{2} \right)  
                                       \sin \left[ k j + \frac{\varphi_k}{2} \right]                                      
                                      \end{array} \right]  
\;\;\;\; , 
\label{app.11}
\eneq
\noindent
for the negative-energy solution, with 

\beq
\cos ( \varphi_k ) = - \frac{\frac{J}{2} \cos ( k ) -  h}{\epsilon_k} 
\:\:\: , \:\:
\sin ( \varphi_k ) = \frac{\frac{J}{2} \sin ( k ) }{\epsilon_k}
\:\:\:\: . 
\label{app.12}
\eneq
\noindent
with the allowed values of $k$ again given by Eq.(\ref{app.10bis}). 
On rewriting the dispersion relation as 

\beq
\epsilon_k = \sqrt{\left( \frac{J}{2} \mp h \right)^2 + J h \left[ \cos ( k ) 
\pm 1 \right] }
\:\:\:\:,
\label{app.13}
\eneq
\noindent
we see that the system presents a single-fermion excitation gap $\Delta_w 
= \left| \frac{J}{2} - | h | \right|$, with quantum phase transitions at 
the quantum critical points 
$ \frac{J}{2} = \pm h$ and the gap $\Delta_w$ correspondingly closing at 
$k = \pi$ or at $k = 0$. The junction of quantum-critical Ising chains has been
largely discussed by Tsvelik \cite{tsvelik_Ising,tsvelik_exact}. In the main 
text of this paper we instead focused onto the off-critical 
regime, with a fully gapped JW fermion excitation spectrum for the single chains. 
When the off-critical chain lies in the magnetically ordered phase, corresponding 
to the topological superconducting phase of Kitaev Hamiltonian (that is, 
within the parameter range $ \frac{| 2 h |}{J} < 1$), additional
low-energy sub-gap modes arise, which, in the long-chain limit ($\ell \to \infty$)
evolve into the localized zero-Majorana modes at the endpoints of the chain \cite{kitaev}.
Here, as we are only interested in the boundary physics at the $j=1$-boundary, 
we consider only the solution corresponding to the localized mode near the left-hand 
endpoint of the chain, with  exponentially decaying wavefunction
given by 

\beq
\left[ \begin{array}{c}
        u_j^0 \\ v_j^0
       \end{array} \right]_L = 
       \left[ \frac{\sqrt{J^2 - 4 h^2}}{2 \sqrt{2} h} \right] 
       \: \left[ \begin{array}{c}
                  1 \\ -1 
                 \end{array} \right]
\: \left( \frac{2 h}{J} \right)^j 
\:\:\:\: . 
\label{app.14}
\eneq
\noindent
As expected, the solution in Eq.(\ref{app.14}) becomes non normalizable 
as $\left| \frac{2 h}{J} \right| \geq 1$
and, therefore, it can no more  be accepted as   physically meaningful.


\end{document}